\documentclass[onecolumn,secnumarabic,amsmath,amssymb,prx,superscriptaddress,longbibliography]{revtex4-1}

\usepackage{longtable}
\usepackage{bm}
\usepackage{dcolumn}
\usepackage{epsfig}
\usepackage{graphicx}
\usepackage{subfigure}
\usepackage{comment}
\usepackage{amsfonts}
\usepackage{amsmath}
\usepackage{amssymb}
\usepackage{subfigure}

\usepackage{color}

\usepackage[colorlinks,bookmarks=false,citecolor=blue,linkcolor=blue,urlcolor=blue]{hyperref}

\newcommand{\bs}{\boldsymbol}
\newcommand{\be}{\begin{equation}}
\newcommand{\ee}{\end{equation}}
\newcommand{\bea}{\begin{eqnarray}}
\newcommand{\eea}{\end{eqnarray}}
\newcommand{\RNum}[1]{\uppercase\expandafter{\romannumeral #1\relax}}

\def\bs{\boldsymbol}
\def\vec{\mathbf}
\def\mc{\mathcal}

\usepackage{times}
\usepackage{appendix}

\begin{document}

\title{Strongly frustrated triangular spin lattice emerging from \\
triplet dimer formation in honeycomb Li$_2$IrO$_3$}

\author{Satoshi Nishimoto}
\author{Vamshi M.~Katukuri}
\affiliation{Institute for Theoretical Solid State Physics, IFW Dresden, Helmholtzstrasse 20, 01069 Dresden, Germany}
\author{Viktor Yushankhai}
\affiliation{Joint Institute for Nuclear Research, Joliot-Curie 6, 141980 Dubna, Russia}
\affiliation{Max-Planck-Institut f\"{u}r Physik komplexer Systeme, N\"{o}thnitzer Str.~38, 01187 Dresden, Germany}
\author{Hermann Stoll}
\affiliation{Institute for Theoretical Chemistry, Universit\"{a}t Stuttgart, Pfaffenwaldring 55, 70550 Stuttgart, Germany}
\author{Ulrich K. R\"{o}{\ss}ler}
\author{Liviu Hozoi}
\affiliation{Institute for Theoretical Solid State Physics, IFW Dresden, Helmholtzstrasse 20, 01069 Dresden, Germany}
\author{Ioannis Rousochatzakis}
\affiliation{Institute for Theoretical Solid State Physics, IFW Dresden, Helmholtzstrasse 20, 01069 Dresden, Germany}
\affiliation{Max-Planck-Institut f\"{u}r Physik komplexer Systeme, N\"{o}thnitzer Str.~38, 01187 Dresden, Germany}
\author{Jeroen van den Brink}
\affiliation{Institute for Theoretical Solid State Physics, IFW Dresden, Helmholtzstrasse 20, 01069 Dresden, Germany}
\affiliation{Department of Physics, Technical University Dresden, Helmholtzstrasse.~10, 01069 Dresden, Germany}

\begin{abstract}
Iridium oxides with a honeycomb lattice have been identified as platforms for the much
anticipated Kitaev topological spin liquid: the spin-orbit entangled states of Ir$^{4+}$
in principle generate precisely the required type of anisotropic exchange.
However, other magnetic couplings can drive the system away from the spin-liquid phase.
With this in mind, here we disentangle the different magnetic interactions in Li$_2$IrO$_3$, a
honeycomb iridate with two crystallographically inequivalent sets of adjacent Ir sites. 
Our {\it ab initio} many-body calculations show that, while both Heisenberg and Kitaev
nearest-neighbor couplings are present, on one set of Ir-Ir bonds the former dominates,
resulting in the formation of spin-triplet dimers. 
%
The triplet dimers frame
a strongly frustrated triangular lattice and by exact cluster
diagonalization we show that they remain protected in a wide region of the phase diagram.
\end{abstract}

\date\today
\maketitle

As early as in the 1970s it was suggested that
quantum spins 
in a solid can, instead of ordering in a certain
pattern, form a fluid-type of ground state -- a quantum spin liquid~\cite{Anderson73,Fazekas74}.
Theory predicts a remarkable set of collective phenomena to occur in spin liquids~\cite{Balents10}.
In the honeycomb-lattice Kitaev spin model~\cite{Kit_kitaev_06}, for instance, a spin-liquid state
that has different topological phases with elementary excitations displaying Majorana statistics
has been anticipated.
This has been argued to be relevant for applications in topological quantum computing \cite{Kit_baskaran_07,Kit_nussinov_08,Kit_vidal_08,Kit_tikhonov_11,Kit_nussinov_13}.

The essential feature of the Kitaev model is that there is a different type of spin coupling for
each of the three magnetic bonds originating from a given $S\!=\!1/2$ spin site, 
$KS_i^{x}S_{j}^{x}$, $KS_i^{y}S_{k}^{y}$ and $KS_i^{z}S_{l}^{z}$, where $j$, $k$ and $l$ are $S\!=\!1/2$
nearest neighbors (NN's) of the reference site $i$ and $K$ is the coupling strength.
However, finding materials in which
   the Kitaev spin model and the spin-liquid ground state
are realized has proven to be very challenging~\cite{Balents10}.
In this respect the strongly spin-orbit coupled honeycomb iridates have recently been brought to the fore~\cite{Ir213_KH_jackeli_09,Ir213_KH_chaloupka_10}.
These compounds have the chemical formula $A_2$IrO$_3$, with $A=$ Na or Li,
and contain Ir$^{4+}$ ions in the center of oxygen octahedra that
   form a planar hexagonal network.
Each Ir$^{4+}$ ion has five electrons in the $5d$ shell which the crystal field splits into a $t_{2g}$ and an $e_g$ manifold.
Since the crystal-field splitting is large, the lowest-energy electron configuration is $t_{2g}^5$.
This is equivalent to the $t_{2g}$ shell containing a single hole with spin $S\!=\!1/2$.
However,
   the $t_{2g}^5$ state additionally bears a finite
effective angular moment $L_{\mathrm{eff}}\!=\!1$.
The strong spin-orbit coupling
   for $5d$ electrons 
therefore splits up the $t_{2g}^5$
manifold
into an effective total angular momentum $\mc{J}\!=\!|L_{\mathrm{eff}}\!+\!S|\!=\!3/2$ quartet and a $\mc{J}\!=\!|L_{\mathrm{eff}}\!-\!S|\!=\!1/2$ doublet.
As for the hole the latter is lowest in energy, an effective spin $\mc{J}\!=\!1/2$ doublet (often referred to as a pseudospin $\tilde S$)
   defines to first approximation
the local ground state of the Ir$^{4+}$ ion.

Whereas the formation of such a local $\mc{J}\!=\!1/2$ doublet is well-known for Ir$^{4+}$
ions
inside an undistorted oxygen octahedron \cite{book_abragam_bleaney}, the remarkable insight of Refs.~\onlinecite{Ir213_KH_jackeli_09,Ir213_KH_chaloupka_10} is that when two such octahedra share an edge, the magnetic superexchange (SE) interactions between the
   $\mc{J}\!=\!1/2$ sites
are in principle precisely of Kitaev type.
This observation has made the $A_2$IrO$_3$ honeycomb iridates prime candidate materials in the search for Kitaev spin-liquid ground states. 

Experimentally, however, both Na$_2$IrO$_3$ and Li$_2$IrO$_3$ have been found to order antiferromagnetically below 15 K
\cite{Ir213_jkj2j3_singh_2012,Ir213_LivsNa_cao_13}.
While inelastic neutron scattering \cite{Ir213_choi_2012}, x-ray diffraction~\cite{Ir213_ye_2012} and resonant inelastic x-ray scattering experiments~\cite{Ir213_zigzag_liu_2011} indicate an  antiferromagnetic (AF) zigzag ordering pattern in Na$_2$IrO$_3$, the nature of the AF ground state of Li$_2$IrO$_3$ is to date unknown~\cite{Ir213_jkj2j3_singh_2012,Ir213_LivsNa_cao_13}.
The questions that arise are therefore (i) which magnetic instability preempts the formation of the spin-liquid state and (ii) how close the system remains to a spin-liquid ground state.

\begin{figure}[!b]
\includegraphics[width=0.90\columnwidth]{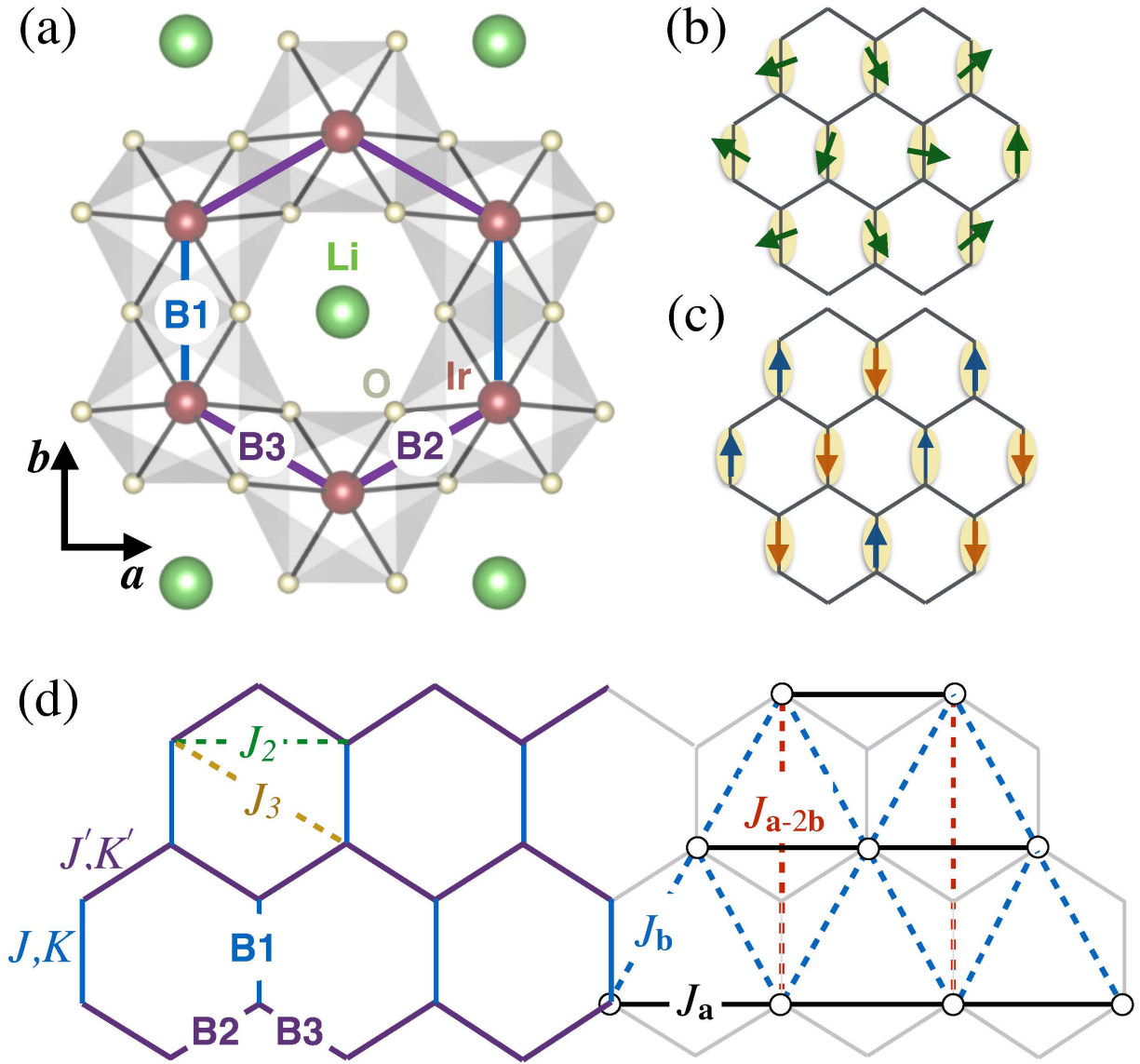}
\caption{ {\bf Honeycomb structure of Li$_2$IrO$_3$ and mapping onto an effective triangular lattice 
of triplet spins.}
(a)
The two distinct sets of NN links \cite{Ir213_omalley_2008} are labeled as 
{B1} (along the crystallographic $b$ axis) and {B2/B3}.
(b) The large FM interaction $J\!=\!-19.2$~meV on {B1} bonds stabilizes rigid $T\!=\!1$
triplets that frame an effective triangular lattice.
The triplet dimers remain protected in a wide region of the phase diagram, including the
incommensurate ICx 
and (c) diagonal-zigzag phase, see text.
(d) Representative exchange couplings for {B1} ($J$,$K$), {B2/B3} ($J'$,\,$K'$), second
neighbor ($J_2$) and third neighbor ($J_3$) paths
on the original hexagonal grid are shown.
$J_{\bs{\delta}}$ ($\bs{\delta}\!\in\!\{\vec{a},\vec{b},\vec{a\!-\!2b}$\})
are isotropic exchange interactions
on the effective triangular net. 
}
\label{structures}
\end{figure}

To answer these fundamental questions it is essential to quantify the relative strengths of the NN magnetic interactions in Li$_2$IrO$_3$, which are already known to be not only of Kitaev, but also of Heisenberg type.
The observed zigzag order in its counterpart system Na$_2$IrO$_3$ has
   indeed been rationalized on the basis
of ferromagnetic (FM) Heisenberg $J$ and AF Kitaev $K$ couplings
\cite{Ir213_KH_chaloupka_12,Ir213_andrade_14,Kee_PRL_2014}
but also interpreted in terms of an AF $J$ and FM $K$
\cite{Ir213_jkj2j3_singh_2012,Ir213_choi_2012,Ir213_trebst_14,Ir213_KH_mazin_2013}.
Recent {\it ab initio} many-body calculations favor the latter scenario, with a relatively 
large FM Kitaev exchange and significantly weaker AF NN Heisenberg interactions in this
material~\cite{Ir213_katukuri_13}.
This scenario is also supported by investigations of model Hamiltonians derived by downfolding
schemes based on density functional theory calculations~\cite{Yamaji_DFT_JK_PRL_2014}.
Besides the NN terms, strongly frustrating longer-range exchange couplings involving the
second ($J_2$) and third ($J_3$) iridium coordination shells were also shown to be relevant
\cite{Ir213_choi_2012,Ir213_jkj2j3_singh_2012,Kee_PRL_2014}, resulting in very rich magnetic
phase diagrams \cite{Ir213_jkj2j3_singh_2012,Ir213_jkj2j3_kimchi_2011,Ir213_katukuri_13}.

Based on the similarity in crystal structure, one might naively expect that the magnetic
interactions
in $A$=Li are similar to the ones in $A$=Na.
Here we show that this is not at all the case.
The strengths of the NN interactions $J$ and $K$ turn out to crucially depend on the Ir-O-Ir
bond angles and distances.
Employing {\it ab initio} wave-function quantum chemistry methods, we find in particular
that in contrast to Na$_2$IrO$_3$ \cite{Ir213_katukuri_13} the Heisenberg coupling $J$ in
Li$_2$IrO$_3$ even has opposite signs for the two crystallographically inequivalent
sets of adjacent Ir sites.
This behavior follows a general trend of $J$ and $K$ as functions of bond angles and interatomic
distances that we have established through a larger, additional set of quantum chemistry
calculations.
The latter show that the NN Heisenberg $J$ has a parabolic dependence on the Ir-O-Ir bond angle
and at around 98$^{\circ}$ changes sign.
This explains why in Na$_2$IrO$_3$, with Ir-O-Ir angles in the range of 98--100$^{\circ}$
\cite{Ir213_choi_2012}, all $J$'s are positive, while in Li$_2$IrO$_3$, which has significantly
smaller bond angles $\sim$$95^{\circ}$~\cite{Ir213_omalley_2008}, the FM component to the
NN Heisenberg exchange is much stronger.
The large FM coupling $J\!\simeq\!-19$~meV on one set of Ir-Ir links in Li$_2$IrO$_3$ 
gives rise to an effective picture of triplet dimers composing a triangular lattice.
To determine the magnetic phase diagram as a function of the strength of the second and third
neighbor exchange interactions ($J_2$ and $J_3$) we use for this effective triplet-dimer model a
semiclassical approach, which we further confront to the magnetic phase diagram for the original 
honeycomb Hamiltonian  calculated by exact cluster diagonalization.
This comparison shows that indeed the triplet dimers act as rigid objects in a wide range of the
$J_2$-$J_3$ parameter space.
We localize Li$_2$IrO$_3$ in a parameter range where the phase diagram has incommensurate magnetic order
 (Coldea 2013)~\cite{Ir213_LivsNa_cao_13,Ir213_jkj2j3_singh_2012}, 
the nature of which goes beyond the standard flat helix modulation scenario, owing to the Kitaev exchange anisotropy. 
\\

\noindent{\bf Results}\\
\noindent{\bf Heisenberg-Kitaev Hamiltonian.}
The experimental data reported in Ref.~\onlinecite{Ir213_omalley_2008} indicate $C_{2h}$ point-group
symmetry for one set of NN IrO$_6$ octahedra, denoted as {B1} in Fig.\,1, and slight distortions
of the Ir$_2$O$_2$ plaquettes that lower the symmetry to $C_{i}$ for the other type of adjacent
octahedra, labeled {B2} and {B3}.
The most general, symmetry-allowed form of the effective spin Hamiltonian for a pair of NN Ir $d^5$ sites, as
discussed in Methods and Supplementary Note 1, is then
\be\label{eqn:Eqn1}
\mc{H}_{\langle ij\rangle\in b}=J_b\,{\tilde {\bf S}}_i\cdot {\tilde {\bf S}}_j +
                                K_b\,{\tilde S}_i^{z_b} {\tilde S}_j^{z_b} +
                                \sum_{\alpha<\beta}\!\Gamma^b_{\!\alpha\beta}
                                ({\tilde S}_i^{\alpha}{\tilde S}_j^{\beta}+{\tilde S}_i^{\beta}{\tilde S}_j^{\alpha})\,.
\ee
The $b$ index refers to the type of Ir-Ir link ($b\!\in$\,\{{B1,B2,B3}\}).
Whereas the Hamiltonians $\mc{H}_{\langle ij\rangle}$ on the Ir-Ir links {B2} and {B3} are related by symmetry, the bond {B1} is distinct
from a symmetry point of view.
Further, ${\tilde {\bf S}}_i$ and ${\tilde {\bf S}}_j$ denote pseudospin-1/2 operators,
$J_b$ is the isotropic Heisenberg interaction and $K_b$ the Kitaev coupling.
The latter plus the off-diagonal coefficients $\Gamma^b_{\!\alpha\beta}$ 
define the symmetric anisotropic exchange tensor.
It is shown below that these $\Gamma^b_{\!\alpha\beta}$ elements are not at all negligible, as assumed in 
the plain Kitaev-Heisenberg Hamiltonian.

In equation (\ref{eqn:Eqn1}), $\alpha$ and $\beta$ stand for components in the local, Kitaev
bond reference frame $\{\vec{x}_b,\vec{y}_b,\vec{z}_b\}$~\cite{Ir213_KH_jackeli_09}.
The $\vec{z}_b$ axis is here perpendicular to the Ir$_2$O$_2$ plaquette (see Methods section, Supplementary Note 2 and Supplementary Figure 1).
In the following, we denote $J_{B1}\!=\!J$, $J_{B2}\!=\!J_{B3}\!=\!J'$, $K_{B1}\!=\!K$, $K_{B2}\!=\!K_{B3}\!=\!K'$
and similarly for the $\Gamma^b_{\!\alpha\beta}$ elements.

\noindent{\bf NN exchange interactions.}
To make reliable predictions for the signs and strengths of the exchange coupling parameters
we rely on many-body quantum chemistry machinery, in particular, multireference configuration-interaction
(MRCI) computations \cite{book_QC_00} on properly embedded clusters.
Multiconfiguration reference wave functions were first generated by complete-active-space self-consistent-field (CASSCF) calculations.
For two NN IrO$_6$ octahedra, the finite set of Slater determinants was defined in the CASSCF treatment in terms of ten electrons and six (Ir $t_{2g}$) orbitals.
The SCF optimization was carried out for an average of the lowest nine singlet and nine triplet states associated with this manifold.
All these states entered the spin-orbit calculations, both at the CASSCF and MRCI levels.
On top of the CASSCF reference, the MRCI expansion additionally includes single and double
excitations from the Ir $t_{2g}$ shells and the $2p$ orbitals of the bridging ligands.
Results in good agreement with the experimental data were recently obtained with this
computational approach for related 5$d^5$ iridates displaying corner-sharing IrO$_6$
octahedra~\cite{Ir113_bogdanov_12, Ir214_katukuri_12,Ir214_katukuri_14}.

Relative energies for the four low-lying states describing the magnetic spectrum of two NN
octahedra and the resulting effective coupling constants are provided in Table~\ref{tab1}.
To derive the latter, we map the quantum chemically computed eigenvalues listed in the table
onto the eigenvalues of the effective magnetic Hamiltonian in equation (\ref{eqn:Eqn1}).
For the effective picture of ${\tilde S}\!=\!1/2$ pseudospins assumed in equation~(\ref{eqn:Eqn1}),
the set of four eigenfunctions contains the singlet
$\Phi^b_{\mathrm{S}}\!=\!(\uparrow\downarrow\!-\!\downarrow\uparrow)/\sqrt2$ and the triplet components
$\Phi^b_{\mathrm{1}}\!=\!(\uparrow\downarrow\!+\!\downarrow\uparrow)/\sqrt2$,
$\Phi^b_{\mathrm{2}}\!=\!(\uparrow\uparrow\!+\!\downarrow\downarrow)/\sqrt2$,
$\Phi^b_{\mathrm{3}}\!=\!(\uparrow\uparrow\!-\!\downarrow\downarrow)/\sqrt2$. 
In $C_{2h}$ symmetry, the ``full" spin-orbit wave functions associated to $\Phi^b_{\mathrm{S}}$,
$\Phi^b_1$, $\Phi^b_2$ and $\Phi^b_3$ transform according to the $A_g$, $B_{u}$, $B_{u}$ and
$A_u$ irreducible representations, respectively.
Since two of the triplet terms may interact, the most compact way to express the eigenstates of
the effective Hamiltonian in equation (\ref{eqn:Eqn1}) is then
$\Psi^b_1\!=\! \Phi^b_1\cos\alpha_b\!+\!i \Phi^b_2\sin\alpha_b$,
$\Psi^b_2\!=\!i\Phi^b_1\sin\alpha_b\!+\!  \Phi^b_2\cos\alpha_b$,
$\Psi^b_3\!=\! \Phi^b_3$, and
$\Psi^b_{\mathrm{S}}\!=\!\Phi^b_{\mathrm{S}}$.
The angle $\alpha_b$ parametrizes the amount of $\Phi^b_1$--$\Phi^b_2$ mixing, related to finite
off-diagonal $\Gamma^b_{\!\alpha\beta}$ couplings.
This degree of admixture is determined by analysis of the full quantum chemistry spin-orbit wave
functions.
The effective parameters provided in Table~\ref{tab1} are obtained for each type of Ir-Ir
link by using the $E^b_1$, $E^b_2$, $E^b_3$, $E^b_{\mathrm S}$ MRCI relative energies and
the $\Phi^b_1$--$\Phi^b_2$ mixing coefficients (see Methods and Supplementary Note 1). For a comparision of the effective parameters derived from CASSCF and MRCI relative energies see Supplementary Tables 1 and 2. 

\begin{table}[!t]
\caption{ {\bf Magnetic spectra of two adjacent Ir$^{4+}$ sites and effective exchange interaction parameters in  Li$_2$IrO$_3$.}
Relative energies of the four low-lying magnetic states and the associated effective exchange couplings (meV)
for each of the two distinct types of [Ir$_2$O$_{10}$] units, {B1} and {B2}/{B3}~\cite{Ir213_omalley_2008}, are shown. 
The energy of the singlet is taken as reference.
Results of spin-orbit MRCI calculations.
}
\label{tab1}
\begin{tabular}{ccc}
\hline
\hline\\[-0.25cm]
Energies \& effective couplings
&{\it b}=B1 \footnote{
$\measuredangle$(Ir-O-Ir)=95.3$^{\circ}$, $d$(Ir-Ir)=2.98 ($\times$2), $d$(Ir-O$_{1,2}$)=2.01 \AA .}
&{\it b}=B2/B3 \footnote{
$\measuredangle$(Ir-O-Ir)=94.7$^{\circ}$, $d$(Ir-Ir)=2.98 ($\times$4), $d$(Ir-O$_{1}$)=2.08, $d$(Ir-O$_{2}$)=1.97 \AA .
O$_1$ and O$_2$ are the two bridging O's.}\\
\hline\\[-0.15cm]
$E_{\mathrm{S}}^b(\Psi_{\mathrm{S}}^b)$ &$0.0$    &$0.0$  \\
$E_1^b(\Psi_1^b)$                       &$-17.1$     &$ 1.3$  \\
$E_2^b(\Psi_2^b)$                       &$-24.8$     &$-3.4$  \\
$E_3^b(\Psi_3^b)$                       &$-21.6$     &$-7.1$ \\[0.20cm]
$J_b$                                   &$-19.2$   &$0.8$  \\
$K_b$                                   &$-6.0$    &$-11.6$\\
$\Gamma_{\!x_by_b}^b$                   &$-1.1$    &$4.2$  \\
$\Gamma_{\!z_bx_b}^b\!=\!
-\Gamma_{\!y_bz_b}^b$                   &$-4.8$    &$-2.0$ \\
\hline
\hline
\end{tabular}
\end{table}

\begin{figure}[!b]
\includegraphics[angle=270,width=0.95\columnwidth]{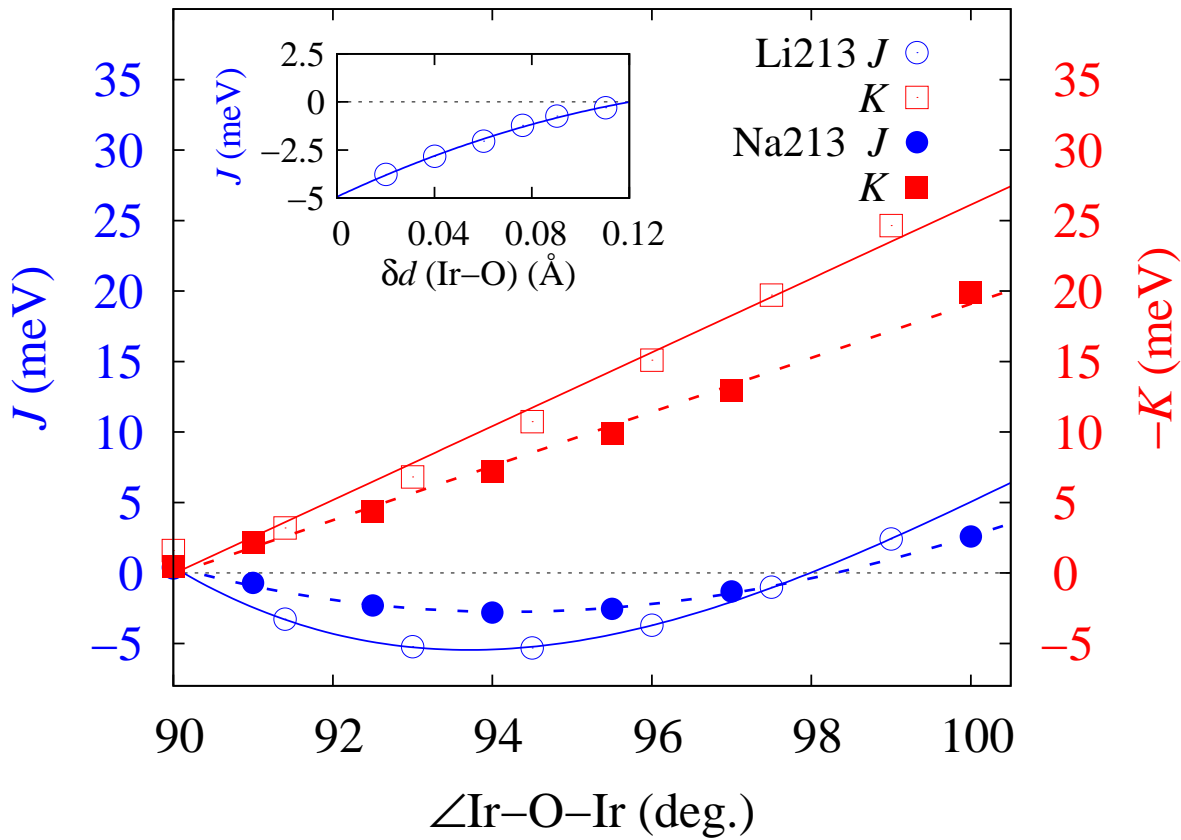}
\caption{ {\bf Variation of the Heisenberg and Kitaev exchange couplings with the Ir-O-Ir angle
in idealized honeycomb structural models.}
Results of spin-orbit MRCI calculations are shown, for NN Ir-Ir links in both Li213 (continuous lines) and Na213 (dashed).
For each system, the NN Ir-Ir distances are set to the average value in the experimental crystal
structure~\cite{Ir213_omalley_2008,Ir213_choi_2012} and the Ir-O bond lengths are all the same.
Consequently, $J\!=\!J'$ and $K\!=\!K'$.
The variation of the Ir-O-Ir angles is the result of gradual trigonal compression.
Note that $|J|$,\,$|K|$\,$\lesssim$\,1 meV at 90$^{\circ}$.
Inset: dependence of the NN $J$ in Li213 when the bridging O's are gradually shifted in
opposite senses parallel to the Ir-Ir axis.
}
\label{JK_NaLi213}
\end{figure}

For the {B1} links in Li$_2$IrO$_3$ (Li213) we find that both $J$ and $K$ are FM, in contrast to Na$_2$IrO$_3$ (Na213), where $J$ is AF for all pairs of Ir NN's \cite{Ir213_katukuri_13}.
Insights into this difference between the Li and Na iridates are
provided by the curves plotted in Fig.~\ref{JK_NaLi213}, displaying the dependence of the NN $J$ on the amount of trigonal distortion for simplified structural models of both Li213 and Na213. The trigonal compression of the O octahedra translates into Ir-O-Ir bond angles larger than 90$^{\circ}$. Additional distortions giving rise to unequal Ir-O bond lengths, see the footnotes in Table~\ref{tab1}, were not considered in these idealized lattice configurations.
Interestingly, we find that for 90$^{\circ}$ bond angle -- the case for which most of the SE models are constructed~\cite{Ir213_KH_jackeli_09,Ir213_KH_chaloupka_10,Ir213_KH_chaloupka_12,Ir213_KH_mazin_2013} -- both $J$ and $K$ are very small, $\lesssim$1\,meV.

In Fig.~\ref{JK_NaLi213}, while $|K|$ monotonously increases with the Ir-O-Ir bond angle, $J$ displays a parabolic behavior and with a minimum at around 94$^{\circ}$.
Indeed on the basis of simplified SE models one expects $J$ to be minimal at around a bond angle close to 90$^{\circ}$.
However, from SE models it is at the same time expected that $K$ is substantial for such bond angles.
The difference between the {\it ab initio} results for 90$^{\circ}$ Ir-O-Ir angles and the predictions of simplified superexchange models originates from assuming in the latter perfectly degenerate Ir $5d$ and O $2p$ orbitals and from the subsequent cancellation of particular inter-site $d-p-d$ exchange paths. The quantum chemistry calculations show that the Ir $5d$ levels are not degenerate (nor the O $2p$ functions at a given site); the symmetry lowering at the Ir/O sites and this degeneracy lifting are related to the strongly anisotropic, layered crystal structure. For the actual honeycomb lattice with trigonal distortions of oxygen cages, one should develop a SE theory using the trigonal $5d$ orbital basis as well as the correspondingly oriented oxygen orbitals. This produces a more general anisotropy than the Kitaev one. This is the essential reason we find at $90^\circ$ for 
Na213 (Ir-Ir average distances of 3.133 \AA):  $J = 0.32, K = - 0.43, \Gamma_{xy} = 2.6, \Gamma_{zx} = -1.3, \Gamma_{yz} = 1.3$
and for 
Li213 (Ir-Ir average distances of 2.980 \AA):    $J = 0.40, K = - 1.60, \Gamma_{xy} = 5.4, \Gamma_{zx} = -2.8, \Gamma_{yz} = 2.8$ meV. 
For both materials $K$ actually turns out to be the smallest of the anisotropic exchange constants at 90$^{\circ}$.
The small value of $K$ may give the impression that only a weak uniaxial anisotropy is active (see Supplementary Table 3).
However, if one diagonalizes the full $\Gamma$ matrix to obtain its principal axes (which in
general are distinct from any crystallographic directions) and corresponding anisotropies,
one finds sizable anisotropic exchange constants as large as few meV.   

Our investigation also shows that the large FM $J$ value obtained for the {B1} Ir-Ir links
in Li213 is the superposition of three different effects (see Fig.~\ref{JK_NaLi213}):
(i) an Ir-O-Ir bond angle smaller than the value of $\approx$98$^{\circ}$ where $J$ changes sign
which in contrast to Na213 takes us into the FM regime,
(ii) the shift to lower values of the minimum of the nearly parabolic $J$ curve in Li213 as 
compared to Na213 and further
(iii) the additional distortions giving rise to three different sets of Ir-O bond lengths
for each IrO$_6$ octahedron.
The latter are significantly stronger in Li213, remove the degeneracy of the Ir $t_{2g}$ levels
and make that the NN {B1} $J$ is even lower than the minimum of the parabola displayed in
Fig.~\ref{JK_NaLi213}.
It is also interesting that the off-diagonal $\Gamma_{yz}$ and $\Gamma_{zx}$ couplings on {B1}
have about the same strength with the Kitaev $K$ (see Table~\ref{tab1}).
Our {\it ab initio} results justify more detailed model Hamiltonian investigations of such
off-diagonal couplings along the lines of Refs.\,\onlinecite{Kee_PRL_2014}, \onlinecite{Ir213_trebst_14}
and \onlinecite{Ir213_katukuri_13}. 

\begin{figure*}[!t]
\includegraphics[width=0.99\linewidth]{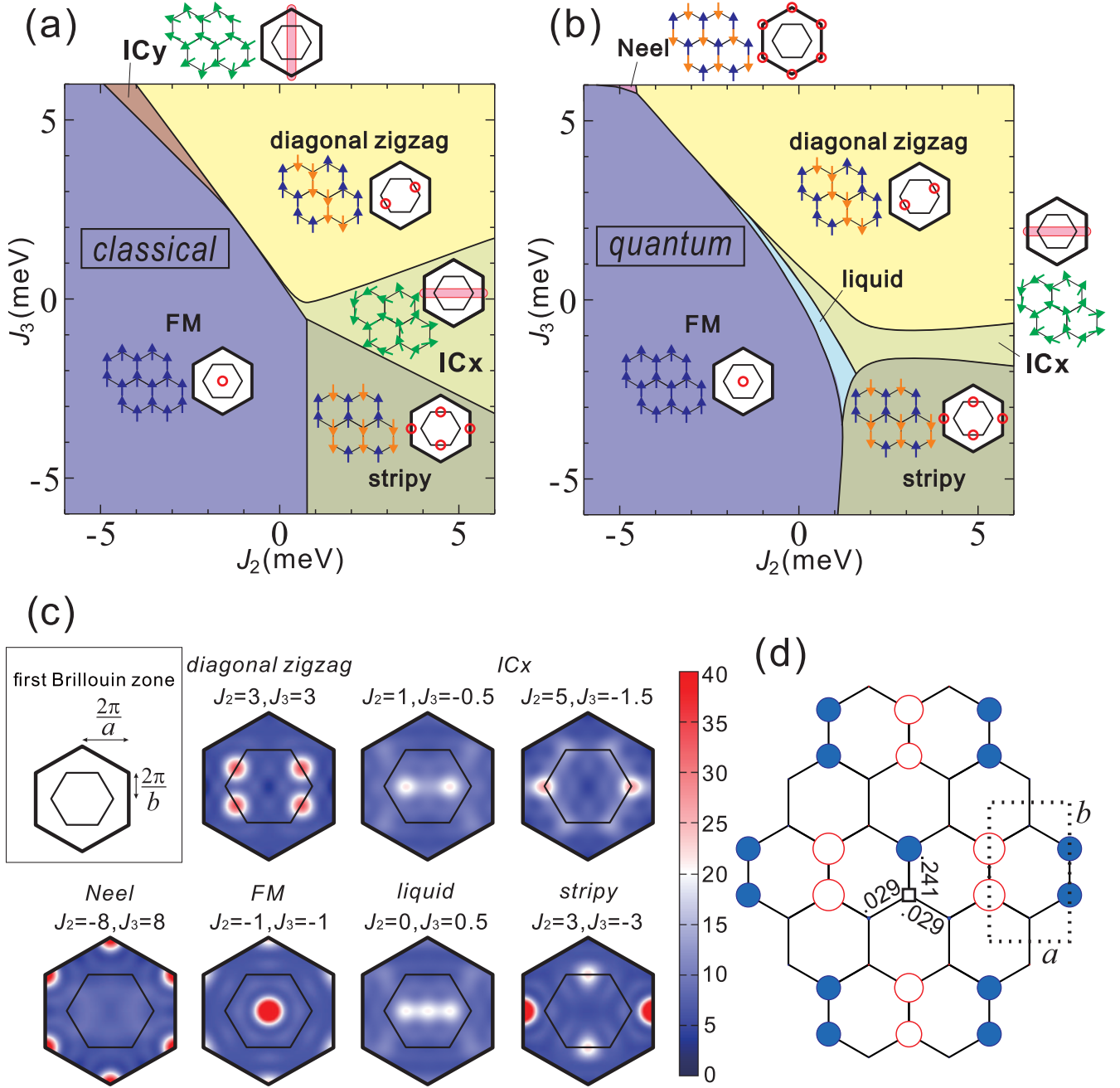}
\caption{ {\bf Magnetic phase diagrams and spin structure factor.}
Phase diagram of Li213 in the $J_2$-$J_3$ plane with the NN couplings listed in Table~\ref{tab1}, along with schematic spin configurations and Bragg peak positions (red circles) for each phase. (a) Classical phase diagram of the effective spin $T\!=\!1$ model on the triangular lattice, found by a numerical minimization of the interaction matrix $\bs{\Lambda}(\vec{k})$ in the BZ. The actual ground-state configurations in the incommensurate regions ICx and ICy can be much richer that the standard coplanar helix states owing to anisotropy, see text.
(b) Quantum mechanical phase diagram for the original spin-1/2 model.
(c) Structure factor $\mc{S}(\vec{k})$ for representative momenta in different phases.
Note that in the ICx phase, the peak position $(\pm Q_a,0)$ takes values between $0< Q_a \le 2\pi/a$, depending on $J_2$ and $J_3$.  (d) Long range spin-spin correlation profiles $\langle \vec{\tilde S}_i\!\cdot\vec{\tilde S}_j\!\rangle$ at $J_2\!=\!J_3\!=\!3$ (i.e., inside the diagonal-zigzag phase), as obtained by ED calculations. The reference Ir site is shown as a black square rectangle, positive (negative) correlations are denoted by filled blue (open red) circles whose radii scale with $|\langle \vec{\tilde S}_i\!\cdot\!\vec{\tilde S}_j\rangle|$. We also show explicitly the actual values for the NN correlations.}\label{phasediagram}
\end{figure*}

For the {B2} and {B3} links, the Ir-O bonds on the Ir-O$_2$-Ir plaquette have different
lengths and the symmetry of the two-octahedra block is lowered to $C_i$ \cite{Ir213_omalley_2008}.
The {\it ab initio} data show that consequently the FM exchange is here disfavored such that
$J'$ turns AF.
This is illustrated in the inset of Fig.~\ref{JK_NaLi213}, where we plot the evolution of the
NN Heisenberg coupling when in addition to trigonal distortions the bridging ligands on the
Ir-O$_2$-Ir plaquette are gradually shifted in opposite senses parallel to the Ir-Ir axis.
For the reference equilateral plaquette, the Ir-O-Ir bond angle is set to the average value
in the experimental structure, $95^{\circ}$ \cite{Ir213_omalley_2008}.
It is seen that such additional distortions indeed enhance the AF contribution to the Heisenberg
SE.
Although the bond symmetry is lower for the {B2}/{B3} links, the analysis of the spin-orbit wave
functions shows however negligible additional mixing effects and the {\it ab initio} results were
still mapped onto a $C_{2h}$ model with $\Gamma^b_{\!z_bx_b}\!=\!-\Gamma^b_{\!y_bz_b}$.

\noindent{\bf Longer range interactions.}
Having established the dominant NN couplings we now turn to the magnetic phase diagram of
Li213 including the effect of second and third neighbor Heisenberg interactions $J_2$ and $J_3$.
The latter are known to be sizable \cite{Ir213_KH_mazin_2013} and to significantly influence
certain properties \cite{Ir213_choi_2012,Ir213_jkj2j3_singh_2012,Ir213_jkj2j3_kimchi_2011,Ir213_katukuri_13}.
However, since correlated quantum chemistry calculations for these longer-range interaction terms
are computationally much too demanding, we investigate their effect by computations for extended effective
Hamiltonians that use the {\it ab initio} NN magnetic interactions listed in Table~\ref{tab1} and adjustable 
isotropic $J_2$, $J_3$ exchange couplings.

\noindent{\bf Triplet dimers.}
With strong FM exchange on the {B1} bonds, a natural description of the system consists in
replacing all {B1} pairs of Ir 1/2 pseudospins by rigid triplet degrees of freedom.
This mapping leads to an effective model of spin $T\!=\!1$ entities on a triangular lattice,
captured by the Hamiltonian 
\bea
\label{eq:Heff}
\mc{H}_{\sf eff} \!=\! \sum_{\vec{R}}( 
\vec{T}_{\vec{R}}\!\cdot\!\bs{\Gamma}_1\!\cdot\!\vec{T}_{\vec{R}} \!+\!
\sum_{\bs{\delta}}(J_{\bs{\delta}} \vec{T}_{\vec{R}}\!\cdot\!\vec{T}_{\vec{R}+\bs{\delta}}
\!+\! \vec{T}_{\vec{R}}\!\cdot\!\bs{\Gamma}_{2,\bs{\delta}}\!\cdot\!\vec{T}_{\vec{R}+\bs{\delta}})),
\eea
where $\bs{\delta}\!\in\!\{\vec{a},\vec{b},\vec{a\!-\!2b}$\} (see Fig.~\ref{structures}(d) and Supplementary Figure 2). It includes both on-site ($\bs{\Gamma}_1$) and intersite ($J_{\bs{\delta}}$,
$\bs{\Gamma}_{2,{\bs{\delta}}}$) effective interaction terms.
While the explicit expressions of these terms are given in Methods, the essential features of
the model are as follows.
First, among the few different contributions to $\bs{\Gamma}_1$, there is an effective coupling
of the form $\frac{K}{2}\left(T_{\vec{R}}^z\right)^2$.
Since $K\!<\!0$, this term selects the two triplet components with $T_z\!=\!\pm 1$ and therefore
acts as an easy-axis anisotropy.
Second, there are two different types of effective exchange couplings between NN triplets,
see Fig.~\ref{structures}\,(d).
This asymmetry reflects the constitutive difference between bonds {B1} and {B2}/{B3}.
Finally, there is also an effective longer-range exchange driven by the $J_3$ interaction in the
original hexagonal model.

According to our {\it ab initio} results, the on-site anisotropy splitting is $|K|/2\simeq 3$\,meV,
about twice the ordering temperature in Li213.
Naively, this may suggest a truncation of the local Hilbert space such that it includes only
the $T_z\!=\!\pm1$ components, which would lead to an effective doublet instead of a triplet description.
However, such a truncation would not properly account for transverse spin fluctuations driven by intersite
exchange (which may even exceed the on-site splitting, depending on the values of $J_2$ and $J_3$) or for
the coupling to the $T_z\!=\!0$ component via off-diagonal terms in $\bs{\Gamma}_1$.
Lacking {\it a priori} a clear separation of energy scales, one is thus left with a description in terms of $T\!=\!1$ triplets.

In momentum space, the effective model takes the form
\be
\mc{H}_{\sf eff}=\sum_{\alpha,\beta,\vec{k}} T_{\vec{k}}^\alpha\cdot\Lambda_{\alpha\beta}(\vec{k})\cdot T_{-\vec{k}}^\beta~,
\ee
where $\vec{T}_\vec{k}=\frac{1}{N}\sum_{\vec{R}} e^{i\vec{k}\cdot\vec{R}}\vec{T}_{\vec{R}}$,
$N$ is the number of {B1} bonds and $\bs{\Lambda}(\vec{k})$ is a symmetric 3$\times$3 matrix
(see Supplementary Note 3). 
Since $T\!=\!1$, the classical limit is expected to yield a rather accurate overall description of
the phase diagram.
The minimum eigenvalue $\lambda_\vec{Q}$ of $\bs{\Lambda}(\vec{k})$ over the Brillouin zone (BZ), 
provides a lower bound for the classical ground-state energy~\cite{LT,Bertaut,Litvin,Kaplan}.
As shown in Fig.~\ref{phasediagram}(a), there exist five different regions for $|J_{2,3}|\!\lesssim\!6$~meV, three with commensurate  (FM, diagonal zigzag and stripy) and two with incommensurate (IC) $\vec{Q}$ (we call them ICx  and ICy, with $\vec{Q}=(q,0)$ and $(0,q)$, respectively). In all commensurate regions, the state $\vec{T}_\vec{R}=e^{i\vec{Q}\cdot\vec{R}}\vec{v}_\vec{Q}$ (where $\vec{v}_\vec{Q}$ is the eigenvector associated with $\lambda_\vec{Q}$), saturates the above lower energy bound and in addition satisfies the spin length constraint $|\vec{T}_\vec{R}|\!=\!1$ for all $\vec{R}$. 
%
%
We note in particular that compared to the more symmetric case of Na213 \cite{Ir213_katukuri_13},
only the diagonal-zigzag configurations are favored in Li213, with FM correlations along the
two diagonal directions of the lattice.
The third, horizontal zigzag configuration is penalized by the strong FM Heisenberg coupling on
the {B1} links.
Correspondingly, we expect Bragg peaks only at two out of the three M points of the BZ, 
namely $\vec{Q}=(\pi,\pm\frac{\pi}{\sqrt{3}})$ (see $\mc{S}({\bf Q})$ in Fig.~\ref{phasediagram}(c) and Supplementary Figure 3). 
Turning to the incommensurate regions ICx and ICy, the minimum eigenvalue $\lambda_\vec{Q}$ is nondegenerate, which implies that one cannot form a flat helical modulation that saturates the low energy bound and satisfies the spin length constraint for all $\vec{R}$. Especially for ICx, which is the most likely candidate for Li213 (see below), this opens the possibility for nontrivial nonplanar modulations of the magnetization. 

\noindent{\bf Exact diagonalization calculations.}
To establish the effect of quantum fluctuations and further test the triplet-dimer picture,
we additionally carried out exact diagonalization (ED) calculations on 24-site clusters for
the original honeycomb spin-1/2 model including the effect of $J_2$ and $J_3$.
Periodic boundary conditions were applied, as in previous studies \cite{Ir213_KH_chaloupka_12,Ir213_katukuri_13}.
We calculated the static spin-structure factor $\mc{S}({\bf Q})=\sum_{ij} \langle {\tilde {\bf S}}_i\!\cdot\!{\tilde {\bf S}}_j \rangle \exp[i {\bf Q}\cdot ({\bf r}_i-{\bf r}_j)]$ as a function of $J_2$ and $J_3$ while fixing the NN magnetic couplings to the ones in Table~\ref{tab1}.
For a given set of $J_2$ and $J_3$ values, the dominant order is determined according to the wave number {\bf Q} = ${\bf Q}_{\mathrm{max}}$  providing a maximum of $\mc{S}({\bf Q})$. The resulting phase diagram is given in Fig.~\ref{phasediagram}(b).
For each phase, the real-space spin configuration and the reciprocal-space Bragg peak positions are shown. In the absence of $J_2$ and $J_3$, the system is in a spin-liquid phase characterized by a structureless $\mc{S}({\bf Q})$ (see Fig.~\ref{phasediagram}(c)) that is adiabatically connected to the Kitaev liquid phase for $-K\!\gg\!J$ \cite{Ir213_KH_jackeli_09}.
By switching on $J_2$ and $J_3$, we recover most of the classical phases of the effective spin-1
model, including the ICx phase, albeit with a smaller stability region due to finite-size effects.
That the 24-site cluster correlations do not show the ICy phase may well be an intrinsic effect,
given that the classical ICy region is very narrow.
We also find an AF N\'eel state region which is now shifted to larger $J_3$'s as compared to
Na213 [\onlinecite{Ir213_katukuri_13}], due to the large negative $J$ on { B1} bonds.

We note that detecting the diagonal-zigzag phase by ED calculations requires large-size
setups of lattice sites.
This is related to the proximity of this phase to the special point $\bs{\Gamma}\!=\!0$ where the model is highly frustrated. Indeed, in this limit the classical ground-state manifold consists of a one-parameter family of states with two sublattices of spins with arbitrary relative orientation angle. This situation is common in various well-known frustrated models, such as the $J_1$--$J_2$ model on the square lattice~\cite{j1j2a,j1j2b,j1j2c}. The lifting of the accidental degeneracy, either by quantum fluctuations or due to a finite $\bs{\Gamma}$ (see Supplementary Note 4, Supplementary Figure 4 and 5), and the associated locking mechanism between the two sublattices involve a very large length scale~\cite{LengthScale1,LengthScale2}. This explains why our exact spin-spin correlation profiles provided in Fig.~\ref{phasediagram}(d) show that the two sublattices are nearly decoupled from each other.

Except for the N\'eel and the spin-liquid phase, all other phases feature rigid triplets on the {B1} bonds. This is shown in Fig.~\ref{phasediagram}(d) for the diagonal-zigzag phase at $J_2\!=\!J_3\!=\!3$, where the NN correlation function on the {B1} bonds, $\langle \vec{S}_i\!\cdot\!\vec{S}_j\rangle\!\simeq\!0.24$, almost saturates to the full spin-triplet value of $1/4$. This shows that the effective triplet picture is quite robust.

\noindent{\bf Comparison to experiment.}
Our result for rigid triplet degrees of freedom finds support in recent fits of the magnetic
susceptibility data, which yield effective moments of 2.22\,$\mu_{\mathrm B}$ for Li213
\cite{Ir213_triplet_hosono_13}, much larger than the value of 1.74\,$\mu_{\mathrm B}$ expected
for an isotropic 1/2 spin system.
Triplet dimerization was earlier suggested to occur in the chain-like compound In$_2$VO$_5$
\cite{in2vo5_triplets_attfield_08}.
FM, quintet dimers were also proposed to form in ZnV$_2$O$_4$ \cite{znv2o4o5_FM_dimers_08}.

Turning finally to the nature of the actual magnetic ground state of Li213, we first note
that the longer-range couplings $J_2$ and $J_3$ are expected to be both AF
\cite{Ir213_choi_2012,Ir213_jkj2j3_singh_2012} and to feature values not larger than 5--6 meV
\cite{Ir213_choi_2012} in honeycomb iridates, which suggests that Li213 orders either with
a diagonal-zigzag or ICx pattern.
Recent magnetic susceptibility and specific heat measurements show indeed that the ground state
is very different from zigzag in Li213 \cite{Ir213_LivsNa_cao_13} while inelastic neutron
scattering data (Coldea 2013) 
indicate clear signatures of incommensurate
Bragg peaks.
These experimental findings provide strong support for the ICx spin configuration.
As explained above, the actual nature of this phase goes beyond the standard flat helical modulations because the latter are penalized by the anisotropic exchange terms in the Hamiltonian.
\\

\noindent {\bf Conclusions}\\
To summarize, we have established a microscopic spin model and zero-temperature phase diagram for the layered honeycomb iridate Li$_2$IrO$_3$, one of the proposed realizations of the spin-1/2 Kitaev-Heisenberg model with strongly spin-orbit coupled Ir$^{4+}$ magnetic ions. {\it Ab initio} quantum chemistry electronic-structure calculations show that, in contrast to Na$_2$IrO$_3$, the structural inequivalence between the two types of Ir-Ir links has a striking influence on the effective spin Hamiltonian, leading in particular to two very different nearest-neighbor superexchange pathways, one weakly antiferromagnetic ($\simeq$1~meV) and another strongly ferromagnetic (--19\,meV). The latter gives rise to rigid spin-1 triplets on a triangular lattice that remain well protected in a large parameter regime of the phase diagram, including a diagonal-zigzag and an incommensurate ICx phase. 
In view of these theoretical findings and of recently reported neutron scattering data (Coldea 2013), 
we conclude that the magnetic ground state of Li$_2$IrO$_3$ lies
in the incommensurate ICx phase.
Settling its detailed nature and properties calls for further, dedicated experimental and
theoretical investigations.
\\

 \ \ \  

\noindent{\bf Methods}\\
\noindent{\bf Embedded-cluster quantum chemistry calculations.} 
All {\it ab initio} calculations were carried out with the quantum chemistry package
{\sc molpro} \cite{MOLPRO_WIREs}.
Embedded clusters consisting of two NN edge-sharing IrO$_6$ octahedra were considered.
To accurately describe the charge distribution at sites in the immediate neighborhood
\cite{CuO2_dd_hozoi_11,NNs_qc_degraaf_99}, the four adjacent Ir$^{4+}$ ions and the closest
22 Li$^{+}$ neighbors were also explicitly included in the actual cluster.
The surrounding solid-state matrix was modeled as a finite array of point charges fitted to
reproduce the crystal Madelung field in the cluster region.
The spin-orbit treatment was carried out according to the procedure described in Ref.~\onlinecite{SOC_molpro},
using spin-orbit pseudopotentials for Ir (see Supplementary Note 1).

Even with trigonal distortions of the oxygen cages, the point-group symmetry of a given block
of two NN IrO$_6$ octahedra is $C_{2h}$.
Since the $C_2$ axis lies here along the Ir-Ir bond, the effective magnetic Hamiltonian for two adjacent
Ir sites is most conveniently expressed in a local reference system $\{\vec{X}_b,\vec{Y}_b,\vec{Z}_b\}$
with $\vec{X_b}$ along the Ir-Ir link ($\vec{Z_b}$ is always perpendicular to the Ir$_2$O$_2$ plaquette).
It reads
\be
\mc{H}_{\langle ij\rangle} = J_b^{(0)}\tilde{\vec{S}}_i\cdot\tilde{\vec{S}}_j
                           + \tilde{\vec{S}}_i\cdot \left(\!\!
\begin{array}{ccc}
A_b  &0    &0 \\
0    &B_b  &C_b \\
0    &C_b  &-A_b-B_b
\end{array}
\!\!\right)
\cdot\tilde{\vec{S}}_j \,,
\label{eq:hamil}
\ee
where $b\!\in$\,\{{B1,B2,B3}\}.
The diagonal elements in the second term on the right hand side sum up to 0 to give a
traceless symmetric anisotropic exchange tensor.
If $X_b$ is $C_{2}$ axis, only one off-diagonal element is nonzero.

In the local Kitaev reference frame \{$\vec{x}_b$,$\vec{y}_b$,$\vec{z}_b$\}, that is rotated
from \{$\vec{X}_b$,$\vec{Y}_b$,$\vec{Z}_b$\} by 45$^{\rm o}$ about the $\vec{Z}_b=\vec{z}_b$
axis (see Supplementary Note 2, Supplementary Figure 1 and Refs.~\onlinecite{Ir213_KH_jackeli_09}, \onlinecite{Ir213_katukuri_13}), the
Hamiltonian shown above in equation (\ref{eq:hamil}) is transformed to the Hamiltonian in equation (\ref{eqn:Eqn1}).
For the latter, the effective exchange couplings are obtained for each type of Ir-Ir link as
\bea
          J_b&=&J_b^{(0)}+\frac{A_b+B_b}{2}\,, \ \            K_b=-\frac{3}{2}(A_b+B_b)\,,\nonumber\\
\Gamma_{xy}^b&=&\frac{A_b-B_b}{2}\,, \quad \qquad   \Gamma_{yz}^b=-\Gamma^b_{zx}=\frac{C_b}{\sqrt{2}}\,,\nonumber
\eea
where the connection to the quantum chemically computed eigenvalues provided in Table~\ref{tab1} (and Supplementary Tables 1 and 2) is
 \bea
J^{(0)}_b&=&\frac{1}{3}(E^b_1 + E^b_2 +E^b_3) - E^b_{\mathrm S}\,,\nonumber\\
      A_b&=&\frac{2}{3} (E^b_1+E^b_2)-\frac{4}{3}E^b_3\,,\nonumber\\
      B_b&=&\frac{1}{2}\left[-A_b \pm \frac{2(E^b_1-E^b_2)}{\sqrt{1+\eta_b^2}} \right] \ \mathrm{and}\nonumber\\
      C_b&=&\frac{\eta_b (A_b+2B_b)}{2}\,.
\eea
$E^b_{\mathrm S}$, $E^b_1$, $E^b_2$, $E^b_3$ are the {\it ab initio} eigenvalues, 
$\eta_b\!=\!\frac{2\zeta_b\sqrt{1-\zeta_b^2}}{1-2\zeta_b^2}$ and $\zeta_b\!=\!\sin\alpha_b$ , where $\alpha_b$ is the mixing parameter.

\noindent{\bf Effective spin $T\!=\!1$ description.}
To find the effective interactions between the {B1} triplet dimers, we begin by deriving the
equivalent operators in the $T_{\vec{R}}\!=\!1$ manifold for a {B1} bond at position
$\vec{R}$, where $\vec{T}_{\vec{R}}\!=\!\vec{S}_{\vec{R},1}\!+\!\vec{S}_{\vec{R},2}$ and
$\vec{S}_{\vec{R},1}$, $\vec{S}_{\vec{R},2}$ are the ionic Ir pseudospins defining the {B1}
bond.
If the projector in the $T_{\vec{R}}\!=\!1$ manifold is tagged as $P_T$, we obtain for the dipolar
channel $P_T\vec{S}_{\vec{R},1} P_T\!=\!P_T\vec{S}_{\vec{R},2}P_T\!=\!\frac{1}{2}\vec{T}_{\vec{R}}$,
while for the quadrupolar channel
\bea
P_T[S_{\vec{R},1}^\alpha S_{\vec{R},2}^\beta \!+\! S_{\vec{R},1}^\beta S_{\vec{R},2}^\alpha\!-\!\frac{2}{3}(\vec{S}_{\vec{R},1}\!\cdot\!\vec{S}_{\vec{R},2})\delta^{\alpha\beta}]P_T 
\!=\!\xi Q_{\vec{R}}^{\alpha\beta}\,.~~~\nonumber
\eea
$Q_{\vec{R}}^{\alpha\beta}\!=\!T_{\vec{R}}^\alpha T_{\vec{R}}^\beta\!
                            +\!T_{\vec{R}}^\beta T_{\vec{R}}^\alpha\!
                            -\!\frac{4}{3}\delta^{\alpha\beta}$
is here the quadrupolar operator for a spin-1 degree of freedom and $\xi\!=\!1/2$.
Using equivalent operators we then find the first-order effective Hamiltonian
$\mc{H}_{\sf eff}\!=\!P_T\mc{H}P_T$ of equation~(\ref{eq:Heff}).
The only non-zero elements of the symmetric on-site tensor $\bs{\Gamma}_1$ are
$\Gamma_1^{zz}\!=\!\frac{K}{2}$,
$\Gamma_1^{xy}\!=\!\frac{ A-B}{4}$ and
$\Gamma_1^{yz}\!=\!-\Gamma_1^{xz}\!=\!\frac{C}{2\sqrt{2}}$,
while those of ${\Gamma}_{2,\bs{\delta}}$ are
$\Gamma_{2,\vec{b}}^{yy}\!=\!\Gamma_{2,\vec{a}-\vec{b}}^{xx}\!=\!\frac{K'}{4}$,
$\Gamma_{2,\vec{b}}^{xy}\!=\!\Gamma_{2,\vec{a}-\vec{b}}^{xy}\!=\!-\frac{C'}{4\sqrt{2}}$,
$\Gamma_{2,\vec{b}}^{xz}\!=\!-\Gamma_{2,\vec{a}-\vec{b}}^{yz}\!=\!-\frac{A'-B'}{8}$ and
$\Gamma_{2,\vec{b}}^{yz}\!=\!-\Gamma_{2,\vec{a}-\vec{b}}^{xz}\!=\!-\frac{C'}{4\sqrt{2}}$.
Finally, the intersite isotropic exchange interactions are
$J_{\vec{a}}\!=\!(J_2\!+\!J_3)/2$,
$J_{\vec{a}-2\vec{b}}\!=\!J_3/4$,
$J_{\vec{b}}\!=\!J_{\vec{a}-\vec{b}}\!=\!J_2/2\!+\!J'/4$.
We here employed the global coordinate system $\{\vec{x},\vec{y},\vec{z}\}$ corresponding to
the Kitaev-like frame \{$\vec{x}_b$,$\vec{y}_b$,$\vec{z}_b$\} with $b$\,=\,{B1} (see Supplementary Figure 1).
$J'$, $K'$, $A'$, $B'$ and $C'$ are effective coupling constants on the bonds {B2} and
{B3}, as also mentioned in the main text.
We stress that the on-site quadrupolar term $T_{\vec{R}}^z T_{\vec{R}}^z$ scales with $K/2$,
while in the classical treatment of the original spin-1/2 model such a term would scale with $K/4$.
We can trace this back to the value of $\xi\!=\!1/2$ found above, which in the classical treatment
is $\xi_{\sf clas}\!=\!1/4$.
This means that the
quantum mechanical correlations strongly enhance the effect of the
``on-site'' anisotropy term $K$.
The latter favors alignment along the $z$-axis, against the effect of $K'$ which favors alignment
within the $xy$-plane.
This point is further discussed in Supplementary Note 3 and 4, where we compare the classical treatment of the original
spin-1/2 hexagonal model with the effective spin-1 triangular model.
\\

\noindent{\bf Acknowledgements}\\
We thank R.~Coldea, Y.~Singh, N.~A.~Bogdanov and D.~I.~Khomskii for insightful discussions.
The computations were partially performed at the High Performance Computing Center (ZIH) at Technical University Dresden.
Partial financial support from the German Research Foundation (HO-4427 and SFB 1143) is gratefuly acknowledged.
\\

\noindent{\bf Author contributions}\\
V.M.K. carried out the {\it ab initio} calculations and subsequent mapping of the {\it ab initio} 
data onto the effective spin Hamiltonian, with assistance from L.H., H.S., V.Y. and I.R.
S.N. performed the exact diagonalization calculations. 
I.R. performed the triplet-dimer mapping and analysis, with assistance from S.N. and U.K.R.
L.H. and J.v.d.B. designed the project. 
S.N., V.M.K., L.H., I.R. and J.v.d.B. wrote the paper, with contributions from all coauthors.


\end{document}